%% file: EGC_2020.tex
\documentclass[a4paper,french]{rnti}  

\usepackage[T1]{fontenc}
\usepackage[utf8]{inputenc}

\usepackage{url}

\usepackage{graphicx}

\usepackage{algorithm}
\usepackage{algorithmic}

\usepackage{subcaption}

\titrecourt{Représentations lexicales pour la détection 
d’événements dans un flux de tweets}

\nomcourt{B. Mazoyer et al.}

\titre{Représentations lexicales pour la détection non supervisée d'événements dans un flux de tweets : étude sur des corpus français et anglais}

\auteur{Béatrice Mazoyer\affil{1}\affilsep\affil{2},
        Nicolas Hervé\affil{2},
        
        Céline Hudelot\affil{1},
         Julia Cagé\affil{3}}

\affiliation{
    \affil{1}CentraleSupélec (Université Paris-Saclay), MICS, Gif-sur-Yvette, France\\
          {beatrice.mazoyer, celine.hudelot}@centralesupelec.fr,\\
    \affil{2}Institut National de l’Audiovisuel, Bry-sur-Marne, France\\
          nherve@ina.fr\\
    \affil{3}SciencesPo Paris, Département d'économie, Paris, France\\
          julia.cage@sciencespo.fr
 }

\resume{%
Dans cet article, nous nous intéressons aux approches récentes de plongements lexicaux
en vue de les appliquer à la détection automatique d'événements dans un flux de tweets. Nous modélisons cette tâche comme un problème de clustering dynamique. Nos expériences sont menées sur un corpus de tweets en anglais accessible publiquement
ainsi que sur un jeu de données similaire en français annoté par
notre équipe. Nous montrons que les techniques récentes fondées sur des réseaux de neurones profonds (ELMo, Universal Sentence Encoder, BERT, SBERT), bien que prometteuses sur de nombreuses applications, sont peu adaptées pour cette tâche, même sur le
corpus en anglais. Nous expérimentons également différents types de fine-tuning afin d'améliorer les résultats de ces modèles sur les données en français. Nous proposons enfin une analyse fine des résultats obtenus montrant la supériorité des approches traditionnelles de type tf-idf pour ce type de tâche et de corpus.
}

\summary{%
In this work, we evaluate the performance of recent text embeddings
for the automatic detection of events in a stream of tweets. We model this task as a dynamic clustering problem. Our experiments are conducted on a publicly available corpus of tweets in English 
and on a similar dataset in French annotated by our team. We show that recent techniques based on deep neural networks (ELMo, Universal Sentence
Encoder, BERT, SBERT), although promising on many applications, are not very suitable for this task. We also experiment with different types of fine-tuning to improve these results  on French data. Finally, we propose a detailed analysis of the results obtained, showing the superiority of tf-idf approaches for this task.
}

\begin{document}

%
\section{Introduction}

Les recherches récentes en traitement automatique du langage ont permis
 d'atteindre des performances proches des capacités humaines, notamment
  en ce qui concerne la détection de paraphrase ou 
  l'évaluation de la similarité sémantique entre deux phrases\footnote{Voir les résultats obtenus sur le benchmark GLUE :  \url{gluebenchmark.com}}. Cependant, 
  ces avancées, fondées sur l'entraînement de réseaux de neurones sur de 
  très vastes corpus de textes, sont à nuancer. 
  
En effet, malgré des progrès rapides ces dernières années dans l'adaptabilité
des modèles de traitement du langage (GLUE, le benchmark de référence \citep{wang2018glue}, est
constitué de 9 tâches différentes, et les modèles sont évalués en fonction 
de leur performance moyenne sur toutes ces tâches), il reste difficile
d'adapter ces modèles à de nouvelles tâches. Dans cet article, nous 
nous intéressons ainsi à la similarité thématique de phrases\footnote{Plus précisément, nous nous intéressons à la similarité thématique de texte courts, ici des tweets, qui ont la particularité de n'être pas toujours grammaticalement corrects. Pour simplifier, nous considérons les tweets comme des phrases dans cet article.} (savoir si deux phrases 
parlent du même sujet), qui diffère dans certains cas de la similarité 
sémantique (savoir si deux phrases veulent dire la même chose) évaluée dans GLUE. 
Il est vrai que de nouveaux modèles, comme BERT \citep{devlin_2018_bert} sont prévus pour être facilement adaptés sur de nouveaux corpus. Cependant,
 toute transformation de la tâche initiale, même minime, demande 
d'adapter ce réseau en ayant recours
au fine-tuning sur (au minimum) quelques milliers de phrases, 
ce qui implique des heures d'annotation manuelle pour créer un jeu de données 
adapté. Par ailleurs, même sur une tâche strictement identique, les performances 
annoncées dans la littérature ne sont parfaitement reproductibles que sur des 
corpus en anglais.

Enfin, la plupart de ces modèles 
sont conçus pour être utilisés en entrée de systèmes de bout-en-bout
(\textit{end-to-end}). Par exemple, pour calculer un score de similarité
entre phrases avec BERT, il faut traiter chaque couple de phrases au lieu de
chaque phrase. Ces architectures s'appliquent mal à des systèmes de recherche d'information qui supposent de comparer des centaines de milliers de phrases\footnote{Pour reprendre l'exemple proposé par \citet{reimers_2019_sentence}, 
si l'on cherche à trouver les deux phrases les plus similaires 
dans un corpus de $n$ = 10 000 phrases, le nombre de traitements à 
réaliser est $n(n-1)/2$ = 49 995 000 opérations, ce qui représente 
environ 65 heures de traitement avec BERT sur un GPU V100.}. 
 Pour des 
tâches de clustering ou de recherche d'information, il est plus 
efficace de représenter chaque phrase dans un espace vectoriel où les 
phrases similaires sont proches (ce que l'on appelle un plongement lexical, 
\textit{embedding} en anglais), 
où l'on peut ensuite appliquer des mesures de distance classiques 
(cosinus, distance euclidienne). En utilisant des structures d'index 
adaptées, trouver la paire la plus similaire parmi 10 000 phrases 
peut dès lors s'effectuer en quelques millisecondes \citep{johnson_2019_billion}.

Dans cet article, nous testons différents plongements lexicaux pour une tâche de clustering thématique de tweets : il s'agit de grouper ensemble les documents 
traitant du même sujet, dans des corpus de tweets concernant des événements médiatiques. Nous comparons une représentation traditionnelle des documents sous forme de vecteurs tf-idf \citep{sparck_1972_statistical} à des plongements lexicaux
plus récents : Word2Vec \citep{mikolov2013efficient}, \textit{Universal Sentence
Encoder} \citep{cer2018universal}, SBERT \citep{reimers_2019_sentence}. Afin d'obtenir un plongement lexical à partir de réseaux de neurones profonds, il est possible d'utiliser la sortie d'une couche intermédiaire en tant que représentation vectorielle d'une phrase, ce que nous testons également pour deux modèles, ELMo \citep{peters2018deep} et BERT. 
Nos expériences sont menées sur un jeu de données de 
tweets en anglais accessible publiquement \citep{mcminn2013building} ainsi
 que sur un jeu de données similaire de tweets en français annoté par notre
  équipe. Le code de nos expériences est disponible en ligne\footnote{\url{https://github.com/ina-foss/twembeddings}}. Nous montrons que les plongements lexicaux créés à partir de techniques 
  récentes fondées sur des réseaux de neurones profonds ne permettent pas d'améliorer la performance de l'algorithme de clustering, même sur le jeu de données en anglais. Nous réalisons également plusieurs expériences de fine-tuning afin d'améliorer BERT et SBERT sur le 
  corpus français. Dans ce cadre, nous montrons qu'utiliser un dataset traduit 
  automatiquement de l'anglais en français (ici le benchmark STS\footnote{\url{http://ixa2.si.ehu.es/stswiki/index.php/STSbenchmark}}) est une piste valable pour obtenir
  des jeux de données de fine-tuning en français.
  Enfin, nous proposons une analyse détaillée des résultats obtenus et montrons la supériorité des approches classiques de type tf-idf pour le clustering de tweets. 

\section{État de l'art}
Nous présentons tout d'abord dans cette partie les travaux antérieurs consacrés au clustering thématique de tweets, avec une attention particulière portée aux types de représentations vectorielles utilisées, puis nous détaillons les techniques existantes de plongement lexical de phrases.

	\subsection{Clustering thématique de tweets}
Le clustering thématique de tweets vise à regrouper ensemble les tweets traitant du même sujet. Pour ce type de clustering, il est souvent fait appel à des algorithmes prenant en compte à la fois la similarité thématique des documents et leur proximité temporelle, afin de ne pas regrouper dans le même cluster des tweets émis à des époques très différentes. Par ailleurs, le nombre de thématiques (et donc de clusters) n'est pas connu à l'avance dans la plupart des cas. De ce fait, des techniques telles que l'algorithme ``First Story Detection", utilisé dans le système UMass \citep{allan_2000_detections}, sont souvent appliquées aux tweets. C'est notamment le cas chez \citet{petrovic_streaming_2010}, \citet{hasan_twitternews_2016}, et \citet{repp_extracting_2018}. \citet{petrovic_streaming_2010} proposent une méthode fondée
sur du \textit{Locality Sensitive Hashing} pour accélérer la recherche, tandis que \citet{repp_extracting_2018} introduisent des mini-batches. Cet algorithme "FSD" est détaillé ci-dessous car c'est également celui que nous utilisons.

\label{FSD}
\begin{algorithm}
\caption{``First Story Detection"}
\begin{algorithmic}[1]
\REQUIRE threshold $t$, window size $w$, corpus $C$ of documents in chronological order
\ENSURE thread ids for each document
\STATE $T \leftarrow \left[ \right] ; i \leftarrow 0 $
\WHILE{document $d$ in $C$}
\IF{$T$ is empty}
\STATE $thread\_id(d) \leftarrow i$
\STATE $i \leftarrow i+1$
\ELSE
\STATE $d_{nearest} \leftarrow $ nearest neighbor of $d$ in $T$
\IF{$cosine(d, d_{nearest}) < t$}
\STATE $thread\_id(d) \leftarrow thread\_id(d_{nearest})$
\ELSE
\STATE $thread\_id(d) \leftarrow i$
\STATE $i \leftarrow i+1$
\ENDIF
\ENDIF
\IF{$|T| \geq w$}
\STATE remove first document from $T$
\ENDIF
\STATE add $d$ to $T$
\ENDWHILE
\end{algorithmic}
\end{algorithm}

\citet{sankaranarayanan_twitterstand_2009} et \citet{becker2011beyond} utilisent un autre algorithme fondé sur la distance (pondérée par un facteur temporel) entre la moyenne des vecteurs de chaque cluster et chaque nouveau tweet. \citet{hasan_twitternews_2016} utilisent l'algorithme FSD dans un premier temps pour déterminer le caractère ``unique" d'un tweet (c'est-à-dire s'il est suffisamment éloigné des tweets précédents), mais  l'attribution d'un tweet à un cluster se fait ensuite en fonction de la distance à la moyenne, de façon similaire à \citet{sankaranarayanan_twitterstand_2009}.

Ici, nous utilisons l'algorithme FSD car il est fondé sur une simple mesure de distance, ce qui permet de bien tester la qualité des plongements sémantiques pour des tâches de clustering. Nous introduisons cependant la même variante ``mini-batch" que \cite{repp_extracting_2018} pour diminuer le temps de traitement.

Dans ces travaux, les tweets sont représentés sous la forme de vecteurs tf-idf dans la grande majorité des cas \citep{sankaranarayanan_twitterstand_2009,petrovic_streaming_2010,becker2011beyond, hasan_twitternews_2016}. \citet{repp_extracting_2018} testent différents types de représentations des tweets (moyenne de Word2Vec, moyenne de GloVe, Doc2Vec, moyenne de Word2Vec pondérée par l'idf de chaque mot). Cependant ces représentations sont uniquement testées sur une tâche de classification, et la meilleure représentation (la moyenne de Word2Vec) est ensuite utilisée pour le clustering. Il nous paraît donc important de mettre à jour ces travaux en testant des plongements lexicaux récents, et notamment ceux développés en vue de la représentation de phrases.

	\subsection{Plongements lexicaux de texte}
	
	La méthode de ``vectorisation" la plus couramment utilisée jusque dans les années 2010
était le tf-idf, introduit par \citet{sparck_1972_statistical}. Il s'agit d'une amélioration du principe des vecteurs ``sac de mots" (\textit{``bag of words"}) \citep{harris1954distributional}, où chaque document est décrit par le nombre d'occurrences des mots qu'il contient (``term frequency"). La pondération tf-idf pondère chacun des mots inversement proportionnellement au nombre de documents dans lesquels il apparaît.

    \subsubsection{Plongements lexicaux de mots}

Avec la publication de Word2Vec \citep{mikolov2013efficient} et GloVe \citep{pennington2014glove}, des méthodes fondées sur la prédiction du contexte de chaque mot (ou la prédiction de chaque mot en fonction de son contexte) ont permis de créer des vecteurs de mots porteurs d'une forme de sémantique
autre que leurs fréquence dans le corpus. Ces représentations perdent cependant la faculté
de décrire chaque document par un seul vecteur. Pour contourner ce problème, on représente souvent chaque document par la moyenne des vecteurs des mots qu'il contient.

Avec ELMo \citep{peters2018deep} apparaît une nouvelle génération de modèles,
permettant une représentation des mots non seulement en fonction de leur contexte général (les mots avec lesquels ils sont fréquemment employés dans le corpus d'entraînement), mais aussi en fonction de leur contexte local (dans une phrase en particulier). ELMo est fondé 
sur un réseau de neurones LSTM bi-directionnel entraîné à prédire dans les deux sens le prochain mot d'une séquence (c'est-à-dire prédire le prochain mot d'une phrase, mais aussi, étant donné la fin d'une phrase, prédire le mot venant juste avant). ELMo n'est toutefois pas prévu pour produire des plongements lexicaux de phrases, mais pour être
utilisé en entrée de modèles neuronaux spécifiques à certaines tâches. Les auteurs testent néanmoins la performance de vecteurs de mots directement issus de la première couche ou de la deuxième couche de leur modèle (qui en contient trois) pour une tâche de désambiguïsation par recherche du premier plus proche voisin. Les résultats obtenus sont proches de l'état de l'art.

BERT \citep{devlin_2018_bert} est plus générique encore qu'ELMo, car ce modèle ne nécessite aucune architecture spécifique à chaque type de tâche : il peut être
fine-tuné sur un nouveau jeu de données en ajoutant simplement une couche de sortie. BERT est construit avec une architecture de type \textit{Transformer} \citep{vaswani2017attention}, et pré-entraîné sur deux types de tâches: prédire des mots masqués dans une phrase et prédire la phrase suivante dans un texte. Comme pour ELMo, les auteurs de BERT ne prévoient pas l'extraction de vecteurs de phrases à partir de leur modèle, mais ils démontrent qu'un simple transfer learning (extraction de vecteurs de mots sans fine-tuning utilisés à l'entrée d'un nouveau modèle) permet d'égaler l'état de l'art pour une tâche de détection d'entités nommées.

\subsubsection{Plongements lexicaux de phrases}
\label{embedding_phrases}
Il existe un grand nombre de travaux cherchant à représenter les phrases par des vecteurs génériques, utilisables dans une très grande variété de tâches, notamment pour du transfer-learning. 
Ainsi Skip-Thought \citep{kiros2015skip} est fondée sur une architecture encodeur-décodeur entraînée à générer le passage encadrant une phrase donnée dans un texte. \citet{conneau2017supervised} montrent avec InferSent, un réseau LSTM bi-directionnel siamois (c'est-à-dire que le réseau prend deux phrases en entrée, mais ce sont les mêmes poids qui sont appliqués dans les deux parties du réseau), qu'un apprentissage supervisé fournit de meilleurs résultats pour la création de vecteurs de phrases génériques. En l'occurrence, InferSent est entraîné sur le jeu de données SNLI, qui contient 570 000 paires de phrases en anglais annotées manuellement en trois catégories : la première phrase implique la deuxième, la première phrase contredit la deuxième, ou la première phrase et la deuxième phrase sont mutuellement neutres. \citet{cer2018universal} (\textit{Universal Sentence Encoder}) appliquent les résultats de \citet{kiros2015skip} et \citet{conneau2017supervised} en entraînant une architecture de type \textit{Transformer} à la fois sur des tâches non-supervisées, comme Skip-Thought et sur le jeu de données SNLI, comme InferSent. 

Sentence-BERT \citep[SBERT,][]{reimers_2019_sentence} propose non pas des vecteurs universels, mais une architecture de fine-tuning du modèle BERT spécifiquement adaptée pour produire des plongements lexicaux de phrases adaptés à  certains types de tâches. Ce modèle modifie BERT pour en faire un réseau siamois, complété par une dernière couche dépendante du type de tâche sur lequel le réseau est entraîné. Les auteurs testent leurs représentations sur le jeu de données STS (8628 paires de phrases auxquelles est associé un score de similarité entre 0 et 5) en calculant un simple score de similarité cosinus entre les vecteurs associés à chaque phrase. Ils montrent que les meilleures performances sur le jeu de données STS sont obtenues par un premier fine-tuning sur SNLI puis un second sur le jeu d'entraînement de STS.

Nous détaillons dans la partie suivante les méthodes utilisées pour tester différents types de représentations (tf-idf, Word2Vec, ELMo, BERT, Universal Sentence Embedding, SBERT) pour la tâche de clustering dynamique de tweets.

\section{Méthodes d'évaluation}

Évaluer la ``qualité" d'un plongement lexical pour la représentation de tweets peut se faire selon différentes approches : d'une part, évaluer si la représentation permet une bonne séparabilité des différentes classes (événements). D'autre part, s'assurer que les vecteurs produits se prêtent bien à des mesures de distance, qui sont utilisées pour le clustering. Enfin, il faut évaluer la qualité des modèles pré-entraînés pour différentes langues. Nous avons donc, dans un premier temps, ramené la tâche de détection non supervisée d'événements à une tâche de classification. Dans un second temps, nous avons modélisé le problème de détection d'événements de façon plus réaliste comme un clustering dynamique, en utilisant l'algorithme FSD. Chaque type d'évaluation a été réalisé sur le jeu de données en anglais \citep{mcminn2013building} et sur notre propre corpus en français.

Pour la classification des tweets, nous utilisons un classifieur de type SVM avec un kernel dit ``triangulaire" \citep{fleuret2003scale}. Ce kernel est de la forme $k(x, y) = 1 - ||x - y||$. Nos expériences montrent que ce type de kernel, en plus d'être invariant aux changements d'échelles \citep{fleuret2003scale}, s'applique à la fois à des vecteurs denses et creux, sans modification de paramètres, et obtient des performances similaires aux noyaux paramétriques sur du clustering de texte. Le classifieur est entraîné sur un échantillon aléatoire de 50\% du corpus. La classification est évaluée par la moyenne macro du score F1 de chaque classe.

Pour le clustering, nous utilisons l'algorithme FSD (voir \ref{FSD}) en introduisant des ``mini-batch" de 8 tweets de façon à paralléliser la recherche de plus proche voisin. Les paramètres de cet algorithme sont $w$ (nombre de tweets du passé parmi lesquels on recherche un plus proche voisin) et $t$, le seuil de distance au-dessus duquel un tweet est jugé suffisamment éloigné des tweets passés pour former un nouveau cluster. La valeur de $w$ a été fixée différemment pour chaque corpus : elle est fixée à environ un jour d'historique de tweets, en fonction du nombre moyen de tweets par jour dans chaque corpus. On a ensuite testé différentes valeurs de $t$ pour chaque type de plongement lexical. D'une manière générale, des valeurs de $t$ plus basses entraînent une création de clusters plus fréquente, et donc une meilleure homogénéité intra-cluster, mais peuvent augmenter le sur-clustering. 

La performance du clustering est évaluée par une mesure que nous nommons ``F1 du meilleur appariement" (\textit{best matching F1}). Elle est définie par \citet{yang1998study}: on évalue le score F1 de chaque paire entre les clusters (détectés) et les événements (annotés). On apparie alors chaque événement au cluster pour lequel le score F1 est le meilleur. Chaque événement ne peut être associé qu'à un seul cluster. Le ``F1 du meilleur appariement" correspond donc à la moyenne des F1 des couples cluster/événement, une fois l'appariement réalisé.

\section{Expériences réalisées}

Nous présentons dans cette partie nos expériences, en détaillant le contenu des jeux de données utilisés puis le type de plongements lexicaux testés. Enfin nous revenons sur nos tests de fine-tuning visant à améliorer les performances de BERT et S-BERT sur le corpus français.

\subsection{Jeux de données}
\textbf{Corpus en anglais.} Le corpus \textit{Event2012} \citep{mcminn2013building} est le seul jeu de données publiquement accessible en anglais pour la détection d'événements sur Twitter. Il contient plus de 150 000 tweets annotés, au sein d'un corpus total de 120 millions de tweets collectés entre Octobre et Novembre 2012. Chaque tweet est associé à un identifiant d'événement parmi une liste de plus de 500 événements. L'annotation a été réalisée sur Amazon Mechanical Turk. Conformément aux conditions d'utilisation de l'API Twitter, les auteurs ne partagent pas directement le contenu des tweets, mais seulement leurs identifiants. Le corpus datant de 2012, beaucoup de tweets ont été effacés, et nous n'avons pu obtenir (en Mars 2019) que 66,5 millions de tweets du corpus initial (55\%), et seulement 72 790 tweets annotés (72\%).

\textbf{Corpus en français.} Nous avons annoté notre propre jeu de données de tweets en français, \textit{Event2018}, à partir d'un corpus de 40 millions de tweets collectés entre Juillet et Août 2018. L'annotation a été réalisée manuellement par 3 étudiants en Sciences Politiques\footnote{Tous nos remerciements à Liza Héliès, Siegrid Henry et Antoine Moutiez pour leur annotation attentive}. Les événements ont été tirés aléatoirement parmi les articles parus pendant cette période dans 6 quotidiens français (\textit{Le Monde}, \textit{Le Figaro}, \textit{Les \'Echos} , \textit{Lib\'eration}, \textit{L'Humanit\'e}, \textit{M\'ediapart}) ainsi que parmi les événements particulièrement relayés sur Twitter durant cette période. Les annotateurs avaient pour consigne de trouver pour chaque événement des mots-clefs associés. Pour chaque mot-clef, les tweets contenant cette expression dans le corpus étaient affichés, et les annotateurs devaient sélectionner parmi ceux-ci ceux qui étaient en lien avec l'événement. Au total, 316 événements ont été annotés, avec un score inter-annotateurs \citep{randolph_free_2005} de 0,79. Un travail de regroupement des événements sur plusieurs jours a ensuite été effectué (par exemple, tous les rebondissements de l'affaire Benalla ont été regroupés en un seul événement), pour obtenir 243 "macro-événements", que nous utilisons comme vérité de terrain pour les tâches de clustering et de classification. Au total, 95 796 tweets ont été annoté comme liés à l'un de ces événements. Le corpus est accessible en ligne\footnote{\url{https://dataset.ina.fr/corpus}. Veuillez remplir le formulaire en indiquant le nom du jeu de données (Event2018). Conformément aux CGU de Twitter, nous ne fournissons pas le contenu des tweets mais seulement leurs identifiants. Un script permettant d'obtenir le texte des tweets est fourni dans notre dépôt github. En novembre 2019, 81\% des tweets annotés étaient encore disponibles.} à des fins de recherche. 

\subsection{Plongements lexicaux testés}

Afin de mener nos expériences sur les deux corpus, nous avons choisis des modèles entraînés à la fois sur du français et de l'anglais. Cette sous-partie détaille les modèles utilisés.

\textbf{Tf-idf.} Du fait de la brièveté inhérente aux tweets, nous avons simplifié le calcul de tf-idf à un simple calcul d'idf, car il est peu fréquent qu'un terme soit utilisé plusieurs fois dans le même tweet. La forme utilisée pour calculer le poids d'un terme $t$ dans un tweet est donc $idf(t) = 1 + log(n+1/df(t)+1)$, avec $n$ le nombre total de documents dans le corpus et $df(t)$ le nombre de documents du corpus qui contiennent $t$. 
Nous avons distingué deux modes de calcul pour $n$ et $df(t)$: \textbf{tfidf-dataset} désigne la méthode qui ne décompte que les tweets annotés, et \textbf{tfidf-all-tweets} désigne le mode de calcul qui prend en compte tous les tweets du corpus (plusieurs dizaines de millions de tweets) pour obtenir $n$ et $df(t)$. Pour chaque méthode, nous restreignons le vocabulaire avec une liste de \textit{stop-words} et un seuil $df_{min}$, le nombre minimum de tweets qui doivent contenir $t$ pour qu'il soit inclut dans le vocabulaire. Dans toutes nos expériences, $df_{min}=10$. On obtient donc un vocabulaire de près de 330000 mots en anglais et 92000 mots en français pour \textbf{tfidf-all-tweets}, et de 5000 mots en anglais et 9000 mots en français pour \textbf{tfidf-dataset}.

\textbf{Word2Vec.} Nous avons utilisé des modèles pré-entraînés pour l'anglais, et entraînés nos propres modèles français. Pour chaque corpus, nous distinguons \textbf{w2v-twitter}, les modèles entraînés sur des tweets, et \textbf{w2v-news}, les modèles entraînés sur des articles de presse. Pour l'anglais, w2v-twitter est un modèle pré-entraîné publié par \citet{godin2015multimedia}\footnote{\url{github.com/loretoparisi/word2vec-twitter}} (400 dimensions) et w2v-news est un modèle entraîné sur Google News et publié par Google\footnote{\url{code.google.com/archive/p/word2vec/}} (300 dimensions). En français, w2v-twitter a été entraîné avec l'algorithme CBOW sur 370 millions de tweets collectés entre 2018 et 2019, et w2v-news a été entraîné de la même façon sur 1.9 millions de dépêches AFP collectées entre 2011 et 2019. Les deux modèles ont 300 dimensions.
Comme Word2Vec fournit un plongement lexical de mots et non de phrase, la représentation des tweets est obtenue en moyennant les vecteurs de chaque mot. Deux méthodes ont été utilisées pour la moyenne : une moyenne simple, et une moyenne pondérée par l'idf (tfidf-all-tweets).

\textbf{ELMo.} Pour l'anglais, nous avons utilisé le modèle publié sur TensorFlow Hub\footnote{\url{tfhub.dev/google/elmo/2}}. Pour le français, un modèle entraîné sur du français publié par \citet{che2018towards}\footnote{\url{github.com/HIT-SCIR/ELMoForManyLangs}}. Dans chaque cas, nous utilisons la moyenne des trois couches du réseau comme représentation de chaque mot. La représentation d'un tweet est produite en moyennant ces vecteurs (de dimension 1024).

\textbf{BERT.} Google fournit un modèle en anglais et un modèle multilingue\footnote{\url{github.com/google-research/bert}. Modèles: bert-large, uncased et bert-base, multilingual cased}. Afin d'améliorer les performances du modèle multilingue sur des tweets en français, nous avons poursuivi l'entraînement pendant 150 000 étapes sur des tweets collectés en juin 2018. Nous désignons le modèle multilingue simple par \textbf{bert} et le modèle entraîné sur des tweets par \textbf{bert-tweets}. Dans chaque cas, nous avons utilisé l'avant-dernière couche du réseau (de dimension 768) comme plongement lexical, en moyennant les tokens pour obtenir une représentation de tweet.

\textbf{Universal Sentence Encoder.} Les modèles fournis\footnote{\url{tfhub.dev/google/universal-sentence-encoder-large/3}}\textsuperscript{,}\footnote{\url{tfhub.dev/google/universal-sentence-encoder-multilingual-large/1}} (anglais et multilingue) sont prévus pour fournir des plongements lexicaux de phrases, nous avons donc pu les utiliser tels quels. Les vecteurs calculés sont de dimension 512.

\textbf{Sentence BERT.} Les auteurs de SBERT fournissent des modèles pré-entraînés pour l'anglais\footnote{\url{github.com/UKPLab/sentence-transformers}. Modèle: bert-large-nli-stsb-mean-tokens}. Pour le français, nous avons dû réaliser un fine-tuning du modèle BERT multilingue, que nous présentons dans la sous-partie suivante. Les vecteurs obtenus sont de dimension 768.

\subsection{Fine-tuning pour le corpus français}
Le modèle SBERT est spécifiquement entraîné pour fournir des scores de similarité cosinus. Ainsi, sur le corpus STS de similarité sémantique, la fonction de coût est l'erreur quadratique moyenne entre le score de similarité cosinus entre les vecteurs des deux phrases et le score de similarité évalué manuellement dans le jeu de données. Ce type de modèles paraît particulièrement adapté pour notre algorithme de clustering, et en effet, parmi les plongements sémantiques de phrases (\textit{Universal Sentence Embedding} et SBERT) c'est celui qui fournit les meilleurs résultats de clustering en anglais (voir le tableau \ref{clustering}).

Cependant, le modèle pré-entraîné en anglais est fondé sur le fine-tuning de BERT sur des tâches supervisées (voir la partie \ref{embedding_phrases} pour le détail des tâches SNLI et STS), ce qui ne peut pas être réalisé sans corpus en français annoté. Nous avons donc mis en place deux stratégies pour réaliser un fine-tuning du modèle bert-tweets sur des données en français: d'une part nous avons utilisé \textit{Cloud Translation API}\footnote{\url{cloud.google.com/translate/docs/reference/rest/}} dans la limite d'utilisation gratuite pour traduire une partie du dataset STS (nous avons obtenu 2984 paires de phrases en français). D'autre part, nous avons annoté manuellement 500 paires de titres d'articles de presse sélectionnés car ils contenaient des mots-clefs en commun. L'annotation s'est fait sur une échelle de 0 à 5, de la même façon que pour STS. Cependant, au lieu d'indiquer le degré de similarité sémantique entre les phrases, nous avons plutôt cherché à évaluer si les deux titres décrivaient le même événement. Les deux types de fine-tuning (corpus traduit, ou corpus traduit + corpus annoté) sont désignés par \textbf{sbert-tweets-sts-short} et \textbf{sbert-tweets-sts-long}. Les performances des différentes représentations sont décrites dans la prochaine partie.

\begin{table}[ht]
\begin{center}
\input{results_classif.tex}
\caption{Résultats de la classification des tweets en événements pour chaque corpus. La performance est calculée par la moyenne macro de la mesure F1 pour chaque classe. Chaque mesure est réitérée 5 fois avec des initialisations différentes. Chaque cellule indique la moyenne et l'écart-type de ces 5 mesures, en pourcentages.} \label{classif}
\end{center}
\end{table}
\section{Résultats}
D'une manière générale, pour les deux tâches, aucun des modèles testés ne parvient à faire mieux que le modèle tf-idf calculé sur l'ensemble du corpus (tfidf-all-tweets). Cependant, la performance relative des modèles varie selon la langue, et selon le type de tâche.

Les résultats de classification par SVM (voir tableau \ref{classif}) montrent que BERT et ELMo ne fournissent pas des plongements lexicaux facilement séparables. Les modèles prévus pour être utilisés comme plongements lexicaux (Word2Vec, \textit{Universal Sentence Encoder}, SBERT) obtiennent de meilleurs résultats. Sur le corpus français, les résultats de ces modèles sont similaires à ceux des vecteurs tf-idf. Sur le corpus anglais, les vecteurs tf-idf demeurent les mieux adaptés, avec les vecteurs w2v-news pondérés par les poids tf-idf.

Les vecteurs tfidf-all-tweets donnent également les meilleurs résultats pour la tâche de clustering (tableau \ref{clustering}), et ce de façon encore plus nette que pour la classification. Cela s'explique par la forme des vecteurs tf-idf, particulièrement adaptés aux calculs de similarité cosinus, ainsi que par les caractéristiques propres aux événements dans les deux jeux de données : les mêmes termes sont manifestement largement employés parmi les tweets d'un même événement.
Concernant les modèles neuronaux adaptés aux plongements lexicaux de phrases (SBERT, \textit{Universal Sentence Encoder}), ils ne font pas mieux que les modèles w2v-news pondérés par tf-idf. Sur le corpus anglais, on note que le fine-tuning de \textit{Sentence-BERT} sur des corpus de similarité sémantique (sbert-nli-sts) permet de meilleurs résultats que les vecteurs génériques de \textit{Universal Sentence Encoder}. Notre propre fine-tuning de \textit{Sentence-BERT} (sbert-tweets-sts-short et sbert-tweets-sts-long) ne permet pas de surpasser \textit{Universal Sentence Encoder} sur le corpus français. On note cependant que le corpus de similarité thématique (qui ne contient que 500 paires de phrases) permet d'augmenter de 2 points la performance du clustering. Toutefois, ne disposant pas d'un corpus de taille similaire à SNLI, notre fine-tuning ne parvient pas à d'aussi bons résultats que le modèle anglais.
\begin{table}[ht]
\begin{center}
\input{results_clustering.tex}
\caption{Résultats du clustering de tweets par l'algorithme FSD. La performance est calculée en utilisant le score ``F1 du meilleur appariement" et affichée en pourcentages. Pour chaque modèle, on a sélectionné la meilleure valeur de seuil $t$ par tests successifs.} \label{clustering}
\end{center}
\end{table}


\section{Conclusion}

Dans cet article, nous cherchons à sélectionner le meilleur type de plongement lexical pour une tâche de détection non-supervisée d'événements dans un flux de tweets, que nous modélisons par un clustering dynamique. Nous montrons, sur un corpus en anglais et un corpus en français, qu'une représentation des tweets par tf-idf permet d'obtenir les meilleurs résultats par rapport à Word2Vec, BERT, ELMo, \textit{Universal Sentence Encoder} ou \textit{Sentence-BERT}. Nous montrons également qu'un fine-tuning sur un corpus de quelques centaines de paires de phrases annotées selon leur similarité thématique améliore de deux points les résultats de \textit{Sentence-BERT}, sans permettre néanmoins d'égaler les modèles entraînés sur des centaines de milliers d'examples.

\bibliographystyle{rnti}
\bibliography{biblio}

\Eng

\end{document}

%% file: results_classif.tex
\begin{tabular}{|l|c|c|}
\hline
           \textbf{Modèle} & \textbf{Anglais} & \textbf{Français} \\
\hline
                     bert  &  $74.49 \pm0.41$ &   $78.46 \pm0.68$ \\
              bert-tweets  &                - &    $81.77 \pm0.7$ \\
                     elmo  &  $59.81 \pm0.41$ &   $73.59 \pm0.64$ \\
            sbert-nli-sts  &  $80.55 \pm0.33$ &                 - \\
    sbert-tweets-sts-long  &                - &   $\textbf{86.08} \pm0.86$ \\
         tfidf-all-tweets  &   $\textbf{83.5} \pm0.78$ &   $\textbf{87.79} \pm0.58$ \\
            tfidf-dataset  &  $\textbf{83.46} \pm0.72$ &   $\textbf{87.66} \pm0.69$ \\
                      use  &  $80.26 \pm0.38$ &    $\textbf{87.45} \pm0.6$ \\
                 w2v-news  &  $81.35 \pm0.53$ &    $\textbf{86.59} \pm0.8$ \\
    w2v-news tfidf-weights &  $\textbf{82.39} \pm0.64$ &   $\textbf{87.51} \pm0.71$ \\
              w2v-twitter  &  $76.68 \pm0.53$ &   $\textbf{87.01} \pm0.56$ \\
 w2v-twitter tfidf-weights &   $81.2 \pm0.48$ &   $\textbf{87.73} \pm0.56$ \\
\hline
\end{tabular}

%% file: results_clustering.tex
\begin{tabular}{|l|cc|cc|}
\hline
                           &\multicolumn{2}{c}{Anglais}&\multicolumn{2}{|c|}{Français}\\
\cline{2-5}
                   Modèle  & \textit{t} & \textit{F1} & \textit{t} & \textit{F1} \\
\hline
                     bert  &       0.04 &       39.22 &       0.04 &       44.79 \\
              bert-tweets  &          - &           - &       0.02 &       50.02 \\
                     elmo  &       0.08 &       22.48 &        0.2 &       46.08 \\
            sbert-nli-sts  &       0.39 &       58.24 &          - &           - \\
    sbert-tweets-sts-long  &          - &           - &       0.36 &       67.89 \\
   sbert-tweets-sts-short  &          - &           - &       0.38 &       65.71 \\
         tfidf-all-tweets  &       0.75 &\textbf{70.1}&        0.7 &\textbf{78.05}\\
            tfidf-dataset  &       0.65 &       68.07 &        0.7 &       74.39 \\
                      use  &       0.22 &       55.71 &       0.46 &       74.57 \\
                 w2v-news  &        0.3 &       53.99 &       0.25 &       66.34 \\
    w2v-news tfidf-weights &       0.31 &       61.81 &        0.3 &       75.55 \\
              w2v-twitter  &       0.16 &        43.2 &       0.15 &       57.53 \\
 w2v-twitter tfidf-weights &        0.2 &       53.45 &       0.25 &       71.73 \\
\hline
\end{tabular}